# Combined Stellar Structure and Atmosphere Models: Exploratory Results for Wolf–Rayet Stars

D. Schaerer

*Geneva Observatory, CH–1290 Sauverny, Switzerland*

A. de Koter

*NASA/GSCF USRA, Greenbelt, MD 20771, USA*

W. Schmutz

*Institut für Astronomie, ETH Zürich, CH–8092 Zürich, Switzerland*

**Abstract.** In this paper we present Complete Stellar models (CoStar) for massive stars, which treat the stellar interior and atmosphere, including its wind. Particular emphasis is given to Wolf-Rayet stars. We address the question of the effective temperatures of WNE and WC stars. Our first results show a satisfactory agreement between the CoStar models and the simple temperature correction method applied by Schaller et al. (1992). An analyses of the subphotospheric structure of the WR star models shows the importance of metal opacity. This may be essential for understanding the driving mechanism of Wolf-Rayet winds.

## 1. Introduction

The strong stellar winds of Wolf–Rayet (WR) stars hamper a reliable comparison of stellar parameters deduced from observations with those derived from evolutionary models. This presently limits the comparison of stellar evolution results with observations to luminosities and surface abundances. Essentially, there are three reasons: *i)* The definition of the stellar radius is ambiquous. As the photospheres of Wolf–Rayet stars are already formed in the supersonic part of the wind, the density scale height is large and the radius at given optical depth becomes wavelength dependent. *ii)* The density structure of the winds are not well known. Although it is widely thought that the basic driving mechanism of the winds of WR stars is radiation pressure, we, as yet, do not fully understand the wind dynamics of these stars. Therefore, the density structures are uncertain and have to be assumed or determined empirically. *iii)* With present state-of-the-art models, we are still not able to account consistently for all relevant opacity sources.

In the present study, we compare new CoStar evolutionary tracks with previous, simplified temperature corrections using different opacity sources. In all calculations we have defined the stellar radius as the point where $\tau_{\rm Ross} = 2/3$, allowing a *differential* comparison of $T_{\rm eff}$ values. For the exponent of the velocity law, we assume $\beta = 2$ and for the terminal velocity $v_\infty \approx 2600$ kms$^{-1}$. The mass-loss rates are according to Meynet et al. (1993).

In Sect. 2, we briefly describe the CoStar models. Section 3 discusses $T_{\rm eff}$ comparisons, while Sect. 4 focusses on subphotospheric opacity effects, i.e. at $\tau_{\rm Ross} > 1$. The latter effects may very well be crucial for understanding the mass-loss driving mechanism of Wolf-Rayet stars.

## 2. Method

We have built the first complete stellar models for massive stars, which treat the entire star, including the wind. By coupling the Geneva stellar evolution code with the ISA-WIND non-LTE atmosphere code of de Koter et al. (1993), we construct a consistent solution of the stellar structure including a spherically expanding stellar wind, in contrast to the plane parallel grey atmosphere normally used for evolutionary models. A brief description of our stellar models is given in Schaerer et al. (1994), details about the atmosphere are given in Schaerer & Schmutz (1994, hereafter SS).

## 3. HR–diagram of WNE and WC stars

To explore the WR phases, we have calculated the evolutionary track of a 85 $M_\odot$ star with Z = 0.02, covering the main sequence and subsequent WR phases. Figure 1 shows the CoStar evolutionary track (solid line). The model already enters the WNL phase (defined by its H surface abundance) on the MS. He-burning starts at the point of highest luminosity, from whereon the evolution proceeds to lower luminosities due to a high mass loss. The different WR phases are indicated. Between the WR phases the effective temperature $T_{\rm eff}$ jumps because for different phases, different mass loss rates are adopted. For comparison, the dashed line gives the track if $T_{\rm eff}$ is calculated from the simple correction method of Schaller et al. (1992), which takes electron scattering and line opacity into account. The dotted line shows the correction due to electron scattering only.

As can be seen from Fig. 1, the CoStar models give a $T_{\rm eff}$, which is intermediate between a pure electron scattering atmosphere and the correction method of Schaller et al. It is however important to note that in the exploratory calculations presented here, the non–LTE atmosphere consists of pure He in the WNE and WC phases, and line blanketing has been neglected. Thus the effective temperatures in the CoStar model must be viewed as upper limits.

## 4. WR mass loss – a missing opacity problem ?

Despite several decades of research, the mass loss mechanism of WR stars is still not very well understood (e.g. Willis 1991). The essential problem is how to explain the substantial excess of wind over photon momentum. The proposed solutions all rely on an acceleration mechanism originating in the, for continuum radiation, optically thin part of the stellar wind. A notable approach is to make use of multiple-scattering processes (Lucy & Abbott 1993). Although this has yielded important physical insight, serious problems remain (Schmutz 1994). Realizing that for most of the WR stars, the sonic point is located in the optically thick part of the wind, it seems resonable to expect that the wind acceleration mechanism is somehow connected with processes in this part of the wind. In this section we explore opacity effects in this regime, i.e. at $\tau_{\rm Ross} > 1$.

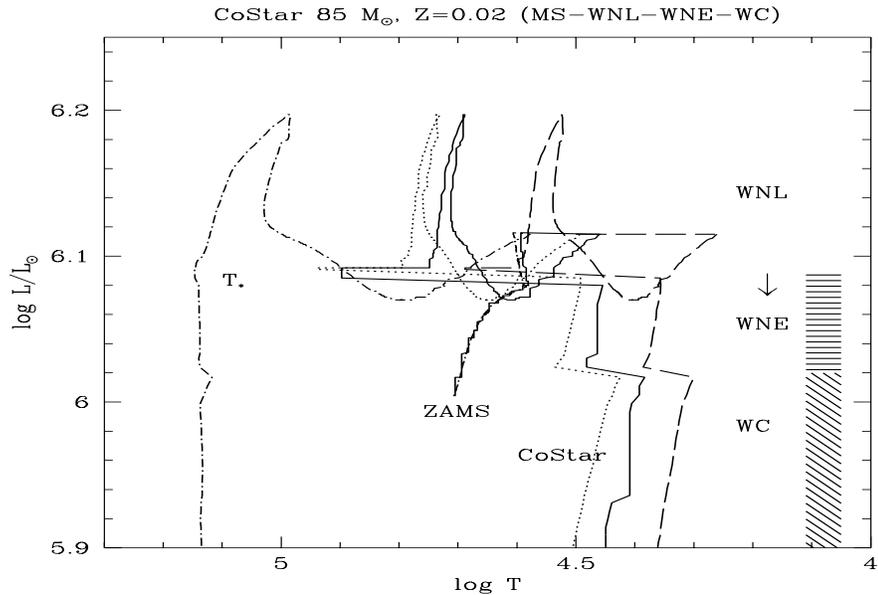

Figure 1. Evolutionary track calculated with our complete stellar models (solid line). Applying the correction method of Schaller et al., which includes an approximate treatment of line opacity, we obtain the dashed track, while the dotted line gives the results with pure electron scattering opacity. The dashed-dotted line shows the track of a wind-free hydrostatic model ($T_\star$ is its effective temperature)

In Figure 2 we show the run of the Rosseland opacity in an atmosphere model of WR 6 (WN5; parameters from Hamann et al. 1993). The inner boundary is at $\tau_{\rm Ross} = 20$. The dashed line shows the opacity of a pure He non–LTE model, while the dotted line gives the opacity obtained from a line blanketed model (see SS). If we adopt the OPAL opacities (Iglesias et al. 1992) at $\tau_{\rm Ross} \geq 1$, we obtain the solid line. The increase of the OPAL opacity represents the low-temperature wing of the opacity peak due to Fe. Admittedly, OPAL opacities are not intended for expanding non–LTE atmospheres. However, we see from the calculated departure coefficients that LTE is roughly valid at $\log \rho > -9.5$. In this region the wind velocity reaches $\approx 300$ km s$^{-1}$. As in an expanding atmosphere $\kappa$ will be increased relative to the static case, the OPAL opacities provide a lower limit for the opacity in the inner part of the wind. Even so, the OPAL data provides roughly twice the opacity of our line blanketed model.

We therefore conclude that in the present treament the opacity calculation needs further improvements at large optical depths. The lack of opacity is probably partly due to continuum opacity of metals, but also to the simplified treatment of the lines in the inner part (see SS), since most important spectral lines are already included in our calculations. For local electron temperatures above $10^5$ K, however, our line list is also inclomplete, since only ionisation stages up to Fe IX are included. Future studies deserve additional effort to include detailed atomic data for higher ionisation stages, such as provided e.g. by the Opacity Project.

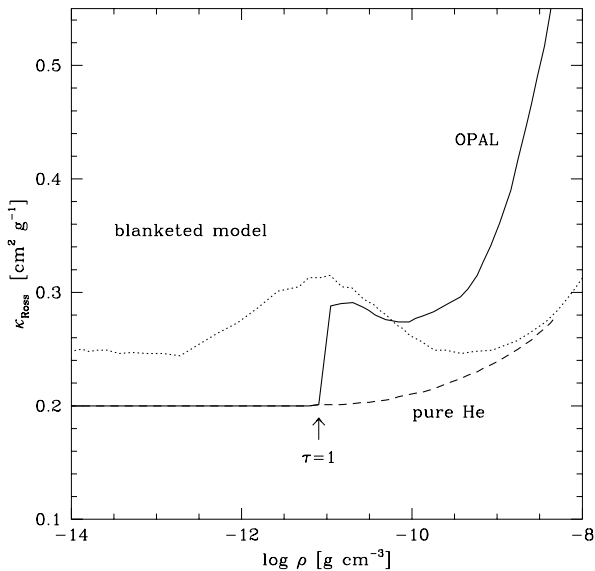

Figure 2. Run of the Rosseland opacity for different models of WR 6. See text for more explanations

At $\tau_{\rm Ross} > 20$, our stellar structure calculations, which use OPAL opacities, show that the opacity peak due to iron, which is located at different depths depending on the effective temperature of the model, causes local supra–Eddington luminosities. According to the treatment of such zones this is simply connected with a small convective zone. However, one could also speculate that this might lead to an optically thick mass outflow (Kato & Iben 1992). In that case, one might expect a dependence of $\dot{M}$ on the Fe abundance, for which there is presently no observational evidence. Nevertheless, the presence of local supra-Eddington luminosities deserves further investigation and may be important for understanding the dynamics of WR wind.